\documentclass{article}

\usepackage[nonatbib,preprint]{neurips_2024}

\usepackage[utf8]{inputenc} 
\usepackage[T1]{fontenc}    
\usepackage{hyperref}       
\usepackage{url}            
\usepackage{booktabs}       
\usepackage{amsfonts}       
\usepackage{nicefrac}       
\usepackage{microtype}      
\usepackage{xcolor}         
\usepackage{cite}
\usepackage{latexsym} 	
\usepackage{amssymb} 
\usepackage{amsbsy}
\usepackage{amsmath}
\usepackage{epsfig}     
\usepackage{graphics,color}
\usepackage{multirow}
\usepackage{hyperref}       

\newcommand{\be}{\begin{equation}}
\newcommand{\ee}{\end{equation}}
\newcommand{\bea}{\begin{eqnarray}}
\newcommand{\eea}{\end{eqnarray}}
\newcommand{\bean}{\begin{eqnarray*}}
\newcommand{\eean}{\end{eqnarray*}}
\newcommand{\bit}{\begin{itemize}}
\newcommand{\eit}{\end{itemize}}

\newcommand{\nn}{\nonumber}
\newcommand{\half}{\frac{1}{2}}

\newcommand{\qqquad}{\qquad\qquad}

\newcommand{\eps}{\epsilon}
\newcommand{\bignorm}{\big|\big|}
\newcommand{\Bignorm}{\Big|\Big|}

\newcommand{\cN}{{\cal N}}
\newcommand{\cO}{{\cal O}}
\newcommand{\E}{\mathbb{E}}
\newcommand{\textt}{\tt}

\usepackage{chngcntr}
\renewcommand{\theequation}{\arabic{section}.\arabic{equation}}
\counterwithin*{equation}{section}
\hypersetup{linkcolor=blue}   

\title{On learning higher-order cumulants in diffusion models}

\author{
Gert Aarts and Diaa E.\ Habibi \\
Department of Physics, Swansea University, Swansea, SA2 8PP,  United Kingdom \\
\textt{g.aarts@swansea.ac.uk} (corresponding author),
\textt{n.e.habibi@swansea.ac.uk}
\And
Lingxiao Wang \\
Interdisciplinary Theoretical and Mathematical Sciences Program (iTHEMS), RIKEN \\ Wako, Saitama 351-0198, Japan \\
\textt{lingxiao.wang@riken.jp}
\And
Kai Zhou \\
School of Science and Engineering, The Chinese University of Hong Kong \\ Shenzhen (CUHK-Shenzhen),
Guangdong, 518172, China  \\
Frankfurt Institute for Advanced Studies, D-60438, Frankfurt am Main, Germany \\
\textt{zhoukai@cuhk.edu.cn} \\
\mbox{} \\
February 28, 2025
}

\begin{document}

\maketitle

\begin{abstract}
To analyse how diffusion models learn correlations beyond Gaussian ones, we study the behaviour of higher-order cumulants, or connected $n$-point functions, under both the forward and backward process. 
We derive explicit expressions for the moment- and cumulant-generating functionals, in terms of the distribution of the initial data and properties of forward process.
It is shown analytically that during the forward process higher-order cumulants are conserved in models without a drift, such as the variance-expanding scheme,  and that therefore the endpoint of the forward process maintains nontrivial correlations. We demonstrate that since these correlations are encoded in the score function, higher-order cumulants are learnt in the backward process, also when starting from a normal prior.
We confirm our analytical results in an exactly solvable toy model with nonzero cumulants and in scalar lattice field theory. 
\end{abstract}



\section{Introduction}

Diffusion models \cite{sohl-dickstein:2015deep,ho:2020denoising,SongErmon2020:generativemodels,Song:2021scorebased} -- see also the review \cite{yang:2022diffusion} -- are a widely used class of deep generative models, able to generate high-quality images and videos via a stochastic denoising process. 
In diffusion models, images are scrambled during the forward process, by applying random noise drawn from a normal distribution to each pixel. It is often stated that at the end of the forward process the images are close to being fully random, that is, without any correlations remaining among pixels. During the forward process, the change in the logarithm of the distribution function is learned (``score matching'')\cite{Hyvarinen:2005jmlr}. In the backward process, this score is applied to initial conditions drawn from a normal distribution and new images are generated (``denoising'') \cite{SongErmon2020:generativemodels}. 
Diffusion models are powerful and widely employed, see e.g.\ Stable Diffusion~\cite{Rombach_2022_CVPR} and DALL-E 2~\cite{2022arXiv220406125R}.

To deepen the understanding of diffusion models, 
it is important to understand how correlations beyond Gaussian ones evolve during both the forward and backward process. To address this, we use generating functionals and lattice field theory as robust and well-understood frameworks. 
On the one hand, lattice field theories are widely studied and well understood in theoretical high-energy physics where they are used to solve strongly interacting quantum field theories describing nature, such Quantum Chromodynamics, see e.g.\ the textbooks \cite{Smit:2002ug,Gattringer:2010zz}. Importantly, interactions between fundamental degrees of freedom are encoded in the higher $n$-point functions, which can be computed using perturbation theory in terms of Feynman diagrams as well as nonperturbatively using lattice simulations. Lattice field theories are therefore useful playgrounds to assess the feasibility of diffusion models (and other generative methods) to learn higher-order $n$-point functions. 
On the other hand, simulations of quantum field theories require the fast generation of ensembles of field configurations and  there is a long history \cite{Creutz:1980zw} of studying strongly interacting quantum field theories numerically by combining the path integral formulation with Monte Carlo methods, after discretisation on a spacetime lattice.
Most standard algorithms, such as Hybrid Monte Carlo (HMC) \cite{Duane:1987de} 
rely on importance sampling, where issues related to critical slowing down remain, see e.g., Ref.\ \cite{Schaefer:2010hu}. It has been suggested -- see e.g.\ the review \cite{Cranmer:2023xbe} -- that methods developed in generative AI can provide an alternative approach to generate ensembles, with configurations playing the role of two- or higher-dimensional images. Indeed, a substantial amount of work has been carried out using normalising flow \cite{rezende2015variational,Noe:2019, Albergo:2019eim,Nicoli:2019gun,Kanwar:2020xzo,Nicoli2021,Nicoli:2023qsl} and variations thereof, such as continuous normalizing flow~\cite{Chen:2018,deHaan:2021erb, Gerdes:2022eve,Caselle:2023mvh} and stochastic normalizing flow \cite{wu2020stochastic, Caselle:2022acb}.
The use of diffusion models to simulate lattice field theories was suggested recently in Refs.~\cite{Wang:2023exq,Wang:2023sry} and extended to U(1) gauge theory in Ref.\ \cite{Zhu:2024kiu}, introducing physics-conditioned diffusion models. To further understand whether diffusion models can be employed in this context, it is of paramount importance to investigate whether higher-order correlations are faithfully reproduced. 
This provides the second motivation for this work, to test the applicability of diffusion models to generate lattice field configurations, via the analysis of higher $n$-point functions. 

We note here that the use of field-theoretical methods to shed light on machine learning methods goes beyond what is sketched above. 
Ref.~\cite{berman2024ncoderquantumfield} proposes to use $n$-point functions to parametrise the latent layer in an autoencoder.
The so-called  neural network/field theory correspondence \cite{Halverson:2020trp,Demirtas:2023fir} exploits the relation between Gaussian processes (or free fields) and neural networks in the limit of infinite width \cite{radford,lee2018deepneuralnetworksgaussian,matthews2018gaussianprocessbehaviourwide,yang2021tensorprogramsiwide,Hashimoto:2024aga}. 
A relation between deep learning and the AdS/CFT correspondence can be found in Ref.~\cite{Hashimoto:2018ftp}.
In quantum field-theoretical machine learning new interaction terms are added on the nodes, to define new systems \cite{Bachtis:2021xoh}. Using the language of field theory may also help in understanding the choice of architecture or hyperparameters, as shown for the Gaussian restricted Boltzmann machine \cite{Aarts:2023uwt}. A correspondence between the dynamics of learning and cosmological expansion can be found in Ref.~\cite{Krippendorf:2022hzj} and the relation between learning and Dyson Brownian motion in Ref.~\cite{Aarts:2024wxi}. The connection between diffusion models and stochastic quantisation \cite{Parisi:1980ys,Damgaard:1987rr,Namiki:1993fd} was pointed out in 
Refs.~\cite{Wang:2023exq,Wang:2023sry}. A further connection with Feynman’s path integral was given in Ref.\ \cite{Hirono:2024zyg}.

The remainder of this paper is organised as follows. In Sec.~\ref{sec:DM} we summarise the basics of diffusion models, using the language of stochastic differential equations.
Moments and cumulants are considered in Sec.~\ref{sec:mom}, where we derive  explicit expressions for the generating functionals. Subsequently we verify the analytical results in the case of an exactly solvable model in Sec.~\ref{sec:toy} and a two-dimensional scalar field theory in Sec.~\ref{sec:lattice}. Conclusions are summarised in Sec.~\ref{sec:con}. App.~\ref{app:DM} contains a brief overview of the numerical implementation of the diffusion model, while App.~\ref{app:Gaussian} gives the analytical solutions for the forward and backward process in the case of a Gaussian target distribution. We consider both score-based, variance-expanding schemes \cite{SongErmon2020:generativemodels,DBLP:journals/corr/abs-2006-09011} and a wide class of denoising diffusion probabilistic models (DDPMs) \cite{ho:2020denoising,sohl-dickstein:2015deep} in the continuous-time limit.

\section{Diffusion models}
\label{sec:DM}

Diffusion models consist of a forward process, in which images or configurations are made more and more noisy, and a backward process, during which new images or configurations are generated in the denoising process. We use the description in term of stochastic differential equations (SDEs) \cite{Song:2021scorebased,song2021maximumlikelihoodtrainingscorebased,karras2022elucidatingdesignspacediffusionbased}, and consider here for notational simplicity one degree of freedom, $x$.\footnote{In Sec.\ \ref{sec:lattice}, we will generalise to this to lattice field theory, using the replacement $x(t)\to \phi(x,t)$.} The forward process is then determined by the stochastic equation,
\be
\dot x(t) = K(x(t),t) + g(t)\eta(t),
\ee
where $K(x(t),t)$ is a possible drift term, $\eta\sim \cN(0,1)$ is Gaussian noise with variance 1, and $g(t)$ is the time-dependent noise strength.  The initial condition for the forward process, $x(0)=x_0$, is determined by the target distribution $P_0(x_0)$, i.e.\ $x_0\sim P_0(x_0)$, which is either known explicitly or implicitly via a data set. We always consider target distributions for which the first moment vanishes, or has been subtracted, $x_0\to x_0-\E_{P_0}[x_0]$.
The forward process runs between $0\leq t\leq T$, where the usual choice is $T=1$. Expectation values are taken by an average over both the noise distribution and the target/data distribution $P_0(x_0)$.

The corresponding backward process is written in terms of $\tau=T-t$, such that $0\leq \tau\leq T$.
In this process the drift should contain a time-dependent term which ensures convergence to the target distribution as $\tau\to T$. This term is given by the change in the logarithm of the distribution $P(x,t)$ in principle. The backward process then reads
\be
\label{eq:bw}
x'(\tau) = -K(x(\tau), T-\tau) + g^2(T-\tau)\partial_x\log P(x,T-\tau) + g(T-\tau)\eta(\tau).
\ee
The initial conditions for the backward process are drawn from a normal distribution with a variance comparable to the final variance of the forward process.   
The second term on the RHS is the essential term to determine, which can be achieved during the forward process via score matching. The numerical implementation of score matching relies on neural networks and is summarised in App.~\ref{app:DM}.   

There is considerable freedom in choosing the drift $K(x,t)$ and the noise strength $g(t)$, including a linear drift, $K(x(t),t) = -\half k(t) x(t)$, 
or no drift term at all, i.e.\ pure diffusion.
A popular choice is the variance-expanding scheme, in which $K(x,t)=0$ and $g(t) = \sigma^{t/T}$, with $\sigma$ a tunable but generically large parameter. The variance at the end of the forward process, $\E[x^2(T)]\sim \sigma^2$, should be substantially larger than the variance of the target distribution. 
In DDPMs the drift is of the form $K(x(t),t) = -\half g^2(t) x(t)$. 
In App.~\ref{app:Gaussian} we give the explicit solutions for a Gaussian target distribution in both schemes, in which both the forward and the backward process can be solved analytically.

\section{Moments and cumulants}
\label{sec:mom}

During the forward process the target distribution evolves to a distribution with a predetermined second moment or variance, while during the backward process the distribution is expected to reverse to the target distribution.  
The question we address here is how higher-order moments or cumulants evolve, both during the forward and the backward process. 
We note here that it is often stated (see e.g.\ Refs.~\cite{Song:2021scorebased,
nakkiran2024stepbystepdiffusionelementarytutorial}) that the final distribution of the forward process approximates a normal distribution, 
but we will see below that this is not the case for pure diffusion.
Knowledge of higher-order cumulants is essential for the application to lattice field theories, since they contain the information on the interactions beyond the free-field limit, as stated in the Introduction.

\subsection{Explicit solution for cumulants}

We start by determining the explicit evolution of the lowest few higher-order cumulants during the forward process. Expressions to all orders, using moment- and cumulant- generating functions, are derived in the next subsection.
 
 We consider the forward process with a linear drift, 
 \be
\dot x(t) = -\half k(t)x(t) + g(t)\eta(t).
\ee
Here the linear coefficient may be time dependent, constant or zero (pure diffusion).
This equation is solved as
\be
\label{eq:solx}
x(t)  = x_0 f(t,0) + \int_0^t ds\, f(t,s)g(s) \eta(s),
\ee
where $x_0\sim P_0(x_0)$ is an initial condition and
\be
\label{eq:fts}
f(t,s) = e^{-\half\int_s^t ds'\, k(s')}.
\ee
Note that for pure diffusion, with $k(t)=0$, $f(t,s)=1$.

We denote the time-dependent moments, or $n$-point functions, as
\be
\mu_n(t) = \E[x^n(t)],
\ee
where the expectation value is taken with respect to the target distribution $P_0$ and the noise distribution. If we only take the expectation with respect to one of these, this will be indicated explicitly with a subscript $P_0$ or $\eta$ respectively. Note that we only consider equal-time  expectation values. Cumulants will be denoted with $\kappa_n(t)$ and are obtained easily using the expansion \cite{cumulants}
\be
\label{eq:expansion}
\kappa_n = \mu_n - \sum_{m=2}^{n-2}\binom{n-1}{m-1}\kappa_m\mu_{n-m},
\ee
with $\mu_1=\kappa_1=0$.

Recall that the target distribution has a vanishing one-point function $\E_{P_0}[x_0]=0$ (after subtraction, $x_0\to x_0- \E_{P_0}[x_0]$, if required). Using the solution (\ref{eq:solx}), the second moment and cumulant then read
\be
\label{eq:mu2}
 \kappa_2(t) = \mu_2(t) = \mu_2(0)f^2(t,0) + \Xi(t),
\ee
where $\mu_2(0)=\E_{P_0}[x_0^2]$ is the variance of the target distribution and
\be
\label{eq:Xi}
\Xi(t) =   \int_0^t ds \int_0^t ds'\,   f(t,s) f(t,s')  g(s) g(s')  \E_\eta[ \eta(s)  \eta(s')]
 =  \int_0^tds\, f^2(t,s)g^2(s).
\ee
The third moment and cumulant are identical and easy to evaluate, 
\be
\kappa_3(t) = \mu_3(t) =  \kappa_3(0)f^3(t,0).
\ee
Using the solution (\ref{eq:solx}), the fourth moment is given by
\be
\mu_4(t) = \mu_4(0)f^4(t,0) + 6\mu_2(0)f^2(t,0)\Xi(t) + 3\Xi^2(t).
\ee
The fourth cumulant is given by
\be
\kappa_4(t) = \mu_4(t) - 3\mu_2^2(t).
\ee
Inserting the expressions for $\mu_4(t)$ and $\mu_2(t)$ then yields 
\be
\kappa_4(t) = \left[ \mu_4(0) - 3\mu_2^2(0) \right] f^4(t,0) = \kappa_4(0) f^4(t,0).
\ee
Similarly, after some algebra, the fifth and sixth cumulants read
\begin{align}
\kappa_5(t) & = \left[ \mu_5(0) -10\mu_3(0)\mu_2(0) \right] f^5(t,0) = \kappa_5(0) f^5(t,0), \\
\kappa_6(t) &= \left[ \mu_6(0) -15\mu_4(0)\mu_2(0) - 10\mu_3^2(0) + 30\mu_2^3(0) \right] f^5(t,0) = \kappa_6(0) f^6(t,0).
\end{align}
Importantly, we find therefore that all cumulants with $n>2$ considered so far take the same form,
\be
\kappa_{n>2}(t) = \kappa_n(0) f^n(t,0),
\ee
i.e.\ they are equal to the product of the $n^{\rm th}$ cumulant of the target distribution and a simple time-dependent function, raised to the power $n$. For pure diffusion, $f(t,0)=1$ and hence
\be
\kappa_{n>2}(t) = \kappa_n(0) \qqquad (\mbox{pure diffusion}),
\ee
i.e.\ the higher-order cumulants are preserved under the forward process. This implies that the final distribution of the forward process is not a normal distribution, but is as correlated as the target distribution, albeit with a different second moment (\ref{eq:mu2}).

A qualitative different result is obtained for a forward process with a nonzero drift term. Taking for simplicity a time-independent one, $k(t)=k$, such that $f(t,0) =\exp(-kt/2)$, one finds that the cumulants decay exponentially, 
\be
\kappa_{n>2}(t) = \kappa_n(0) e^{-\frac{n}{2}kt} \qqquad (\mbox{constant drift}).
\ee
Only the second moment remains nonzero,
\be
 \mu_2(t) = \mu_2(0) e^{-kt} + \int_0^t ds\, e^{-k(t-s)} g^2(s),
\ee
with the dependence on the target data exponentially suppressed. In this case, the final distribution of the forward process is a normal distribution, up to exponentially suppressed terms.

\subsection{Moment- and cumulant-generating functions}

The demonstration above was carried out at a given order. It is helpful to prove these results to all orders, using the moment- and cumulant-generating functions, defined by
\be
Z[J] = \E[e^{J(t)x(t)}], 
\qqquad
W[J] = \log Z[J].
\ee
The normalisation is such that $Z[0]=1$. We take the solution (\ref{eq:solx}) and consider first the average over the noise (see e.g.\ Ref.~\cite{Damgaard:1987rr} for conventions)
\be
Z_\eta[J] = \E_\eta[e^{J(t)x(t)}] 
= \frac{\int D\eta\, e^{-\half \int_0^t ds\, \eta^2(s) + J(t) \left[x_0f(t,0) +\int_0^t ds\, f(t,s)g(s)\eta(s) \right]} }{\int D\eta\, e^{-\half \int_0^tds\, \eta^2(s)} }.
\ee
Completing the square in the exponential and performing the integral over $\eta$ then yields
\be
\label{eq:Zeta}
Z_\eta[J] = e^{J(t) x_0f(t,0) + \half J^2(t) \Xi(t)}.
\ee
Including now the average over the target distribution yields the final expression for the moment-generating function,
\be
Z[J] = \E[e^{J(t)x(t)}] = e^{\half J^2(t) \Xi(t)} \int dx_0\, P_0(x_0) e^{J(t) x_0f(t,0)}.
\ee 
Here we used that the second term in the exponential in Eq.\ (\ref{eq:Zeta}) is independent of $x_0$.
The cumulant-generating function immediately follows as
\be
W[J] = \log Z[J] = \half J^2(t) \Xi(t) + \log \int dx_0\, P_0(x_0) e^{J(t) x_0f(t,0)}.
\ee 
This expression explains the results derived above. The second moment or cumulant is given by
\be 
\kappa_2(t) = \frac{d^2 W[J]}{dJ(t)^2} \Big|_{J=0} = \Xi(t) + \E_{P_0}[x_0^2] f^2(t,0),
\ee
in agreement with Eq.\ (\ref{eq:mu2}). All higher-order cumulants are independent of the stochastic part $\Xi(t)$ and proportional to the cumulants of the target theory, with the replacement $x_0\to x_0f(t,0)$, 
\be
\kappa_{n>2}(t) = \frac{d^n W[J]}{dJ(t)^n} \Big|_{J=0} =  \frac{d^n}{dJ(t)^n} \log \E_{P_0}[ e^{J(t)x_0f(t,0)}]\Big|_{J=0}
= \kappa_n(0)f^n(t,0).
\ee
This is in complete agreement with the explicit results for the fixed-order cumulants derived in the previous subsection.
 
We find therefore that the generating functions have a simple structure. They are of the same form as for the original target distribution $P_0(x_0)$, with the modifications:
\begin{itemize}
\item The degree of freedom, $x_0$, is rescaled with a time-dependent function $f(t,0)$, which results in a multiplication of all connected $n$-point functions with a factor $f^n(t,0$).  
For schemes without a drift (pure diffusion), $f(t,0)=1$, and there is no rescaling.
\item The exception is the two-point function, which contains an additive term $\Xi(t)$, which dominates at the end of the forward process.   
\end{itemize}
This interpretation will be explored further in Sec.\ \ref{sec:lattice}, where we consider the generating functionals for lattice scalar field theory.

\section{Exactly solvable model with nonzero higher-order cumulants}
\label{sec:toy}

To study the dynamics of higher-order cumulants during the forward and backward processes numerically, we consider as target distribution a linear combination of two normal distributions for one degree of freedom, 
\be
\label{eq:p0}
P_0(x) = \half \left[ \cN(x; \mu_0,\sigma_0^2) + \cN(x; -\mu_0,\sigma_0^2) \right].
\ee
This distribution has peaks around $x=\pm\mu_0$ and hence resembles a distribution in a double-well potential, but all moments and hence cumulants can be computed exactly. It is also easy to generate `configurations' numerically, using Gaussian random numbers and shifting them by $\pm\mu_0$ equally often. Finally, as we will see below, the time-dependent score is known analytically as well.  

Since the distribution is even, all odd moments and cumulants are zero. The even moments can be found using the binomial expansion for $(x\pm\mu_0)^{2n}$ and the standard expression for moments of the normal distribution with vanishing mean.
One finds
\be
\mu_{2n} \equiv \E\left[x^{2n}\right]  = \sum_{k=0}^{2n} c_{nk} \sigma_0^{2k}\mu_0^{2n-2k}, 
\qqquad
c_{nk}= \frac{(2n)!}{(2n-2k)!k!2^k}.
\ee
Cumulants are obtained easily using the expansion (\ref{eq:expansion}) or via the cumulant-generating function
\be
\kappa_n = \frac{d^nW[j]}{dj^n}\Big|_{j=0},
\qqquad
W[j] = \log Z[j].
\ee
The moment-generating function $Z[j]$ is a linear combination of the ones for the normal distribution with mean $\pm\mu_0$,  
\be
Z[j] = \E\left[e^{jx}\right] = e^{\half\sigma_0^2j^2}\cosh(\mu_0 j).
\ee
Since 
\be
\label{eq:W}
W[j] = \half\sigma_0^2j^2 + \log\cosh(\mu_0 j),
\ee
only the second cumulant depends on $\sigma_0^2$, and the few lowest cumulants are given by 
\be
\kappa_2  = \mu_0^2+\sigma_0^2, \qquad
\kappa_4  = -2\mu_0^4, \qquad
\kappa_6  = 16\mu_0^6, \qquad
\kappa_8  = -272\mu_0^8.
\ee
We have analysed this model numerically using the variance-expanding scheme and a particular choice of a denoising diffusion probabilistic model.

\subsection{Variance-expanding scheme}

\begin{figure}[bh]
\begin{center}
\includegraphics[width=0.32\textwidth]{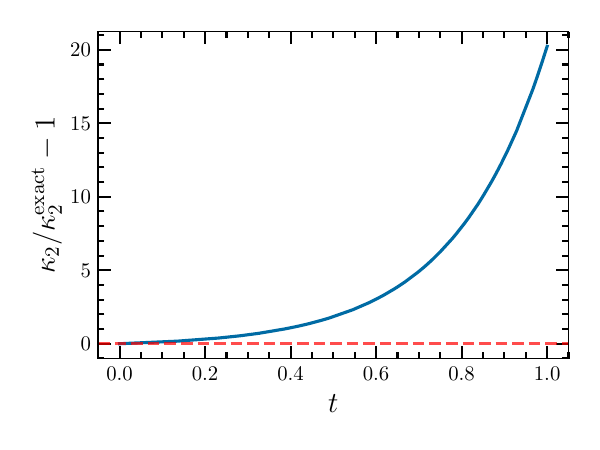}
\includegraphics[width=0.32\textwidth]{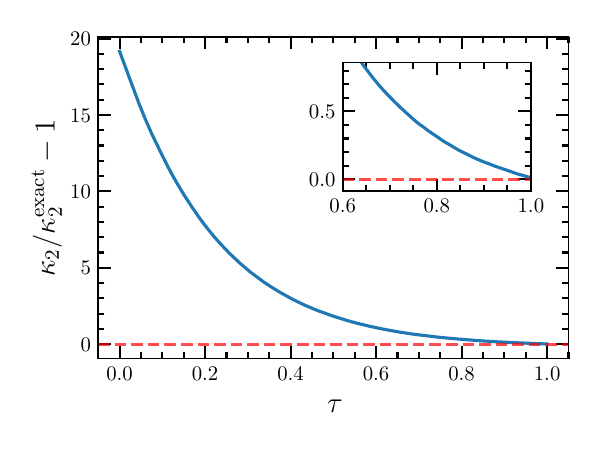}
\includegraphics[width=0.32\textwidth]{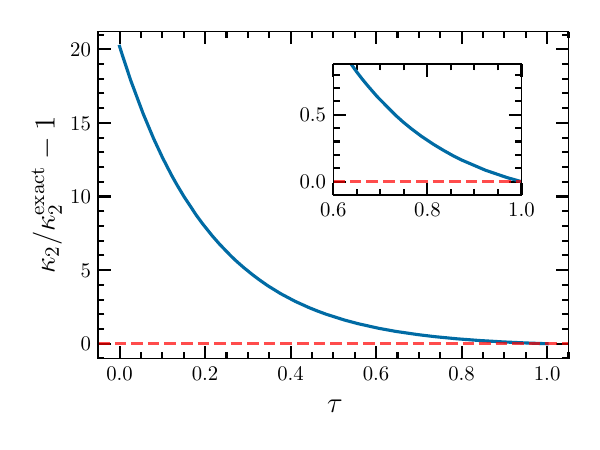} 
\end{center}
  \caption{Evolution of the normalised second moment or cumulant, presented as $\kappa_2/\kappa_2^{\rm exact}-1$, in the two-peak model in the variance-expanding scheme, with $\mu_0=1$ and $\sigma_0=1/4$, during the forward process (left), the backward process with the score determined by the diffusion model (middle), and with the analytical score (right), all using $10^6$ trajectories. 
  The insets zoom in at $0.6<\tau<1$. 
  }
  \label{fig:2peak-2nd-mom}
\end{figure}

We start with the variance-expanding scheme, using $k(t)=0$ and $g(t)=\sigma^{t/T}$, where $\sigma=10$ and $T=1$. The parameters in the target distribution are $\mu_0=1$ and $\sigma_0=1/4$ throughout. 
Some details on the implementation are given in App.\ \ref{app:DM}.
In Fig.~\ref{fig:2peak-2nd-mom} we show the evolution of the second moment (or second cumulant), as $\kappa_2/\kappa_2^{\rm exact}-1$, during the forward and backward process, using $10^6$ trajectories. 
For the latter, we use the score determined by the diffusion model as well as the analytical score (to be discussed below). As expected, the second cumulant increases during the forward process as
\be
\label{eq:kappa2exact}
\kappa_2(t) = \kappa_2^{\rm target} + \frac{T}{\log \sigma^2}\left[ \sigma^{2t/T}-1\right] \sim \frac{T}{\log \sigma^2}\sigma^{2t/T},
\ee
and decreases correspondingly during the backward process. 
The target value is obtained linearly as $\tau\to T$, as shown in the inset, see App.~\ref{app:Gaussian} for details.

\begin{figure}[t]
\begin{center}
\includegraphics[width=0.32\textwidth]{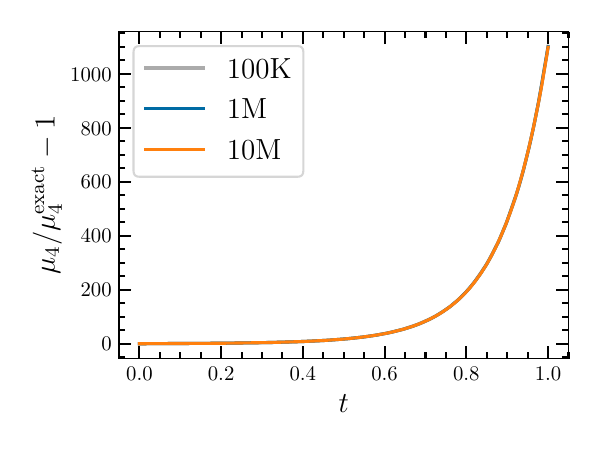} 
\includegraphics[width=0.32\textwidth]{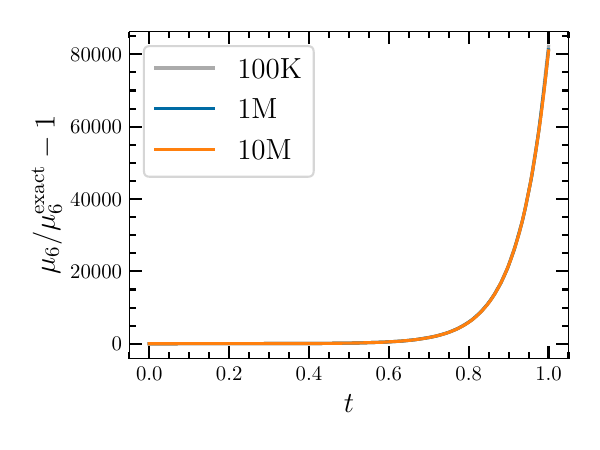}
\includegraphics[width=0.32\textwidth]{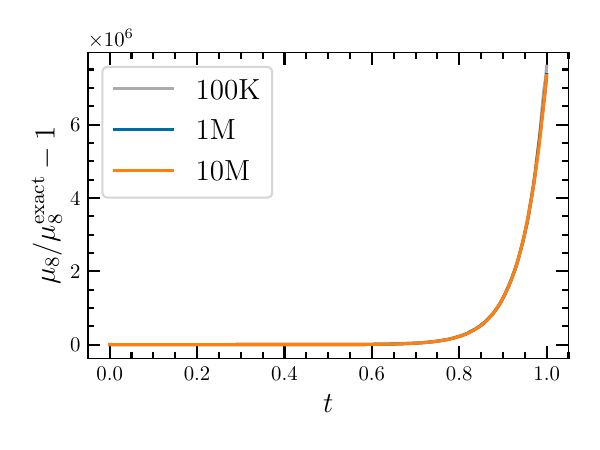} 
\end{center}
\begin{center}
\includegraphics[width=0.32\textwidth]{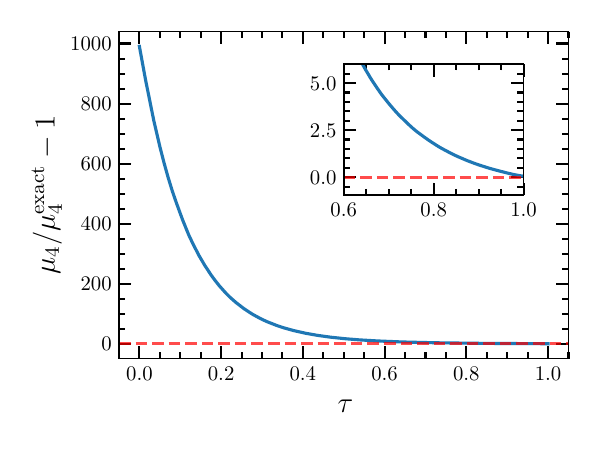} 
\includegraphics[width=0.32\textwidth]{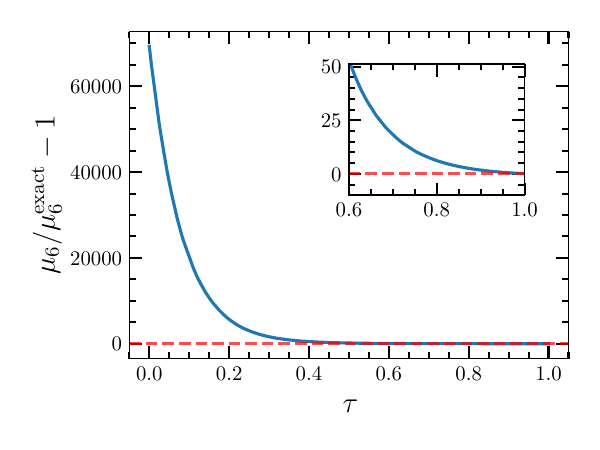}
\includegraphics[width=0.32\textwidth]{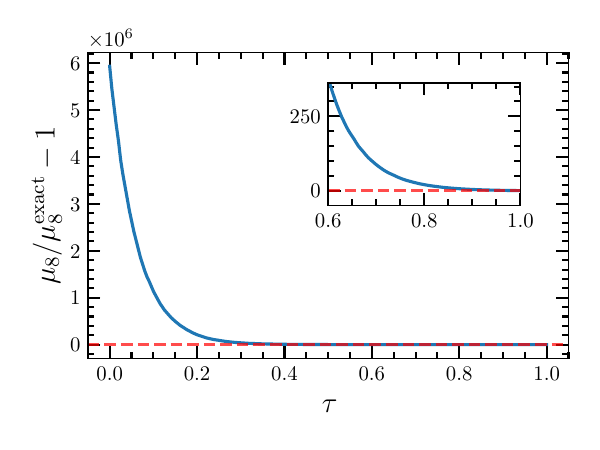} 
\end{center}
  \caption{Evolution of the normalised $4^{\rm th}$ (left), $6^{\rm th}$ (middle) and $8^{\rm th}$ (right) moments, presented as $\mu_n/\mu_n^{\rm exact}-1$, in the two-peak model in the variance-expanding scheme, during the forward process using $10^5, 10^6$ and $10^7$ trajectories (above), and during the backward process with the score determined by the diffusion model, using $10^6$  trajectories (below). Other parameters as above. 
  }
  \label{fig:2peak-bw-mom-DM}
\end{figure}

\begin{figure}[t]
\begin{center}
\includegraphics[width=0.7\textwidth]{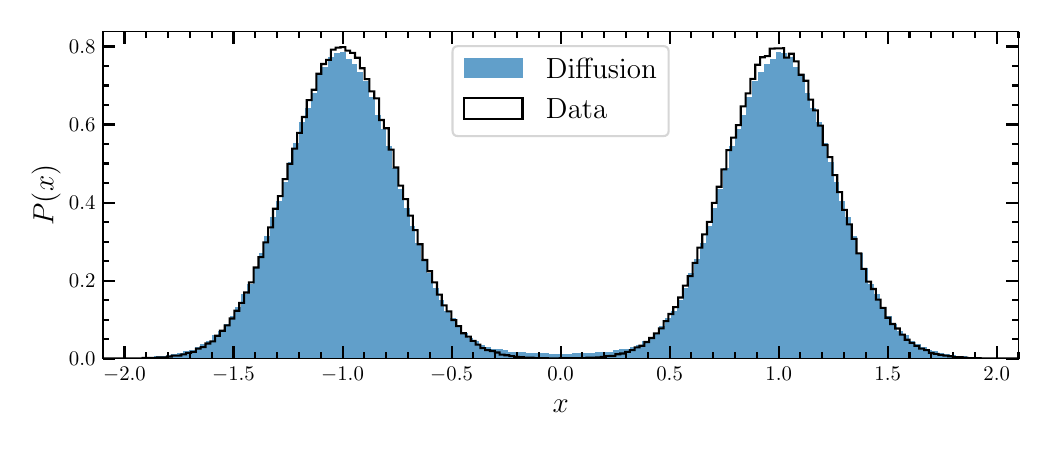}
\end{center}
  \caption{Distribution created by sampling from the target distribution with $\mu_0=1$ and $\sigma_0=1/4$ (Data) and from the trained diffusion model in the variance-expanding scheme (Diffusion), using $10^6$ samples in each case. }
    \label{fig:2peak-dist}
\end{figure}

In Fig.~\ref{fig:2peak-bw-mom-DM} we show the forward and backward evolution of the fourth, sixth and eight moments, as $\mu_n/\mu_n^{\rm exact}-1$. These moments increase (decrease) as $\sigma^{nt/T}$ ($\sigma^{n(\tau-T)/T}$), and hence grow very large. 
For the forward process, we show results for $10^5, 10^6$ and $10^7$ trajectories, which cannot be distinguished. 
The insets indicate that the target values are obtained at the end of the backward process. 
Before turning to the cumulants, we show in Fig.~\ref{fig:2peak-dist} the distribution, as obtained by direct sampling from Eq.~(\ref{eq:p0}) (Data) and as produced by the diffusion model (Diffusion). By eye, the distributions are matching, but we make this precise below.

\begin{figure}[t]
\begin{center}
\includegraphics[width=0.32\textwidth]{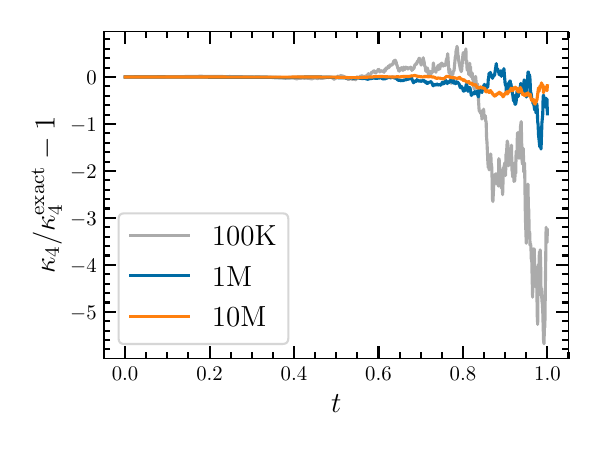} 
\includegraphics[width=0.32\textwidth]{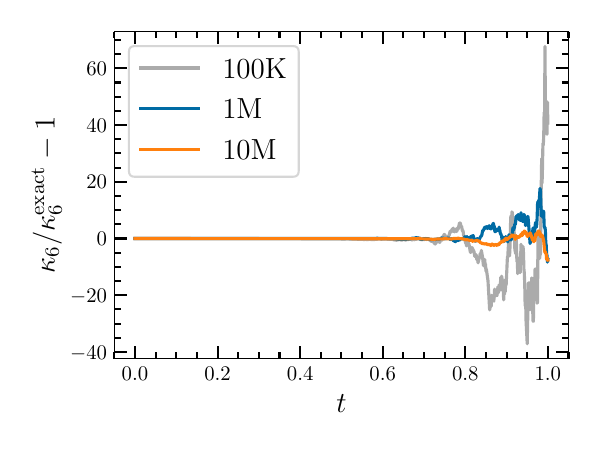}
\includegraphics[width=0.32\textwidth]{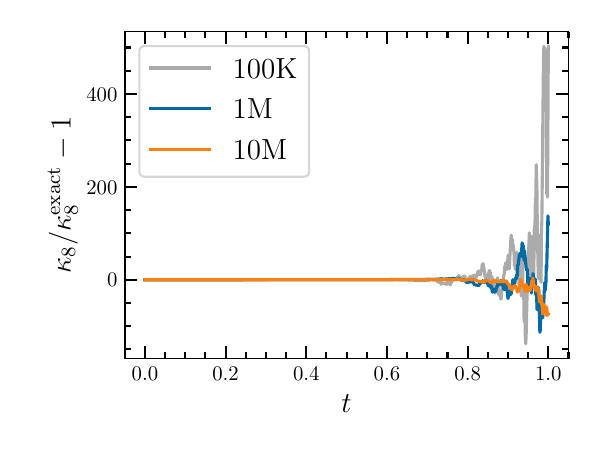} 
\end{center}
\begin{center}
\includegraphics[width=0.32\textwidth]{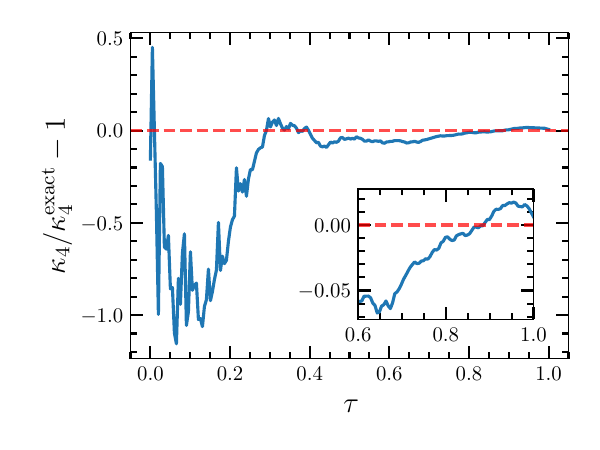} 
\includegraphics[width=0.32\textwidth]{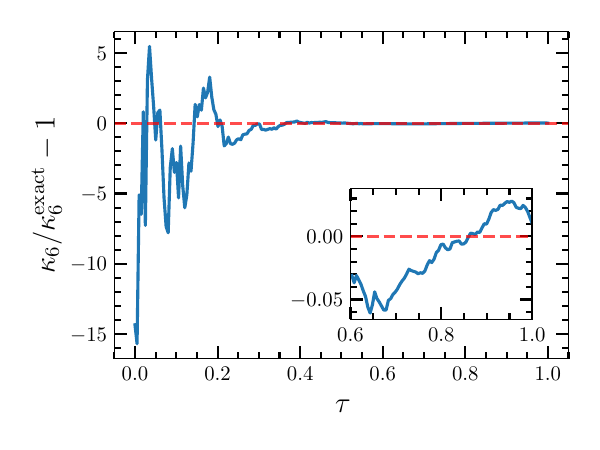}
\includegraphics[width=0.32\textwidth]{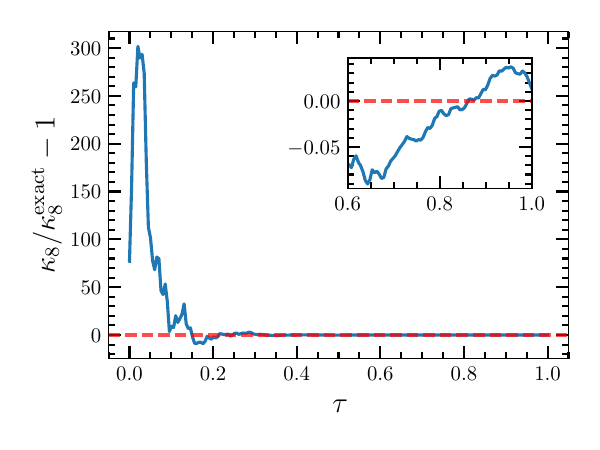} 
\end{center}
  \caption{Evolution of the normalised $4^{\rm th}, 6^{\rm th}$ and $8^{\rm th}$ cumulants, presented as $\kappa_n/\kappa_n^{\rm exact}-1$,
  in the two-peak model in the variance-expanding scheme, during the forward process, using $10^5, 10^6$ and $10^7$  trajectories (above), and during the backward process, with the score determined by the diffusion model, using $10^6$  trajectories (below). Other parameters as above. 
  }
  \label{fig:2peak-bw-cum-DM}
\end{figure}

The forward and backward evolution of the higher moments is dominated by the evolution of the second moment. To study the properties of the distributions in more detail, it is necessary to follow the higher-order cumulants. 
In Fig.~\ref{fig:2peak-bw-cum-DM} (top row) we show the fourth, sixth and eight cumulants, as $\kappa_n/\kappa_n^{\rm exact}-1$, during the forward process. The prediction from the previous section is that these should be conserved during the evolution. 
We observe that they are indeed approximately constant, except towards the end of the forward process. The latter can be understood as the effect of incomplete cancellations. 
Higher-order cumulants are obtained as differences between (higher-order) moments $\mu_n(t)$, which each grow large, as shown in Fig.~\ref{fig:2peak-bw-mom-DM}.
The time evolution hence depends on precise cancellations, which requires sufficient statistics. This is demonstrated by including expectation values with $10^5, 10^6$ and $10^7$  trajectories. The apparent numerical instability is reduced as the number of trajectories increases. This supports the analytical result that the higher-order cumulants are preserved, and that the distribution at the end of the forward process is as correlated as the target distribution (in the limit of an infinite number of trajectories).  

The same effect is observed during the backward process, as shown in Fig.~\ref{fig:2peak-bw-cum-DM} (bottom row), using $10^6$  trajectories. Initial conditions are drawn from a normal distribution and hence the cumulants are expected to be zero initially. 
Here we observe the reverse behaviour. Near the start of the backward process, the second and higher moments are large, leading to only a partial cancellation with a finite number of trajectories. After some time, however, the cumulants become approximately constant and approximately equal to the target value. The fluctuations shown in the inset reflect the stochastic evolution of predominantly the second moment, which approximates the target value as $\tau\to T$, as do the cumulants.

One possibility is that the large fluctuations at the start of the backward process are the result of a poorly learned score. This can be tested in this model, since the time-dependent distribution and hence the score can be obtained analytically. It is then possible to follow the backward process with the exact drift term, using a finite number of trajectories.
The time-dependent distribution during the forward process in the case of pure diffusion reads
\be
P(x,t) = \half \left[ \cN(x; \mu_0,\sigma^2(t)) + \cN(x; -\mu_0,\sigma^2(t)) \right],
\ee
where $\sigma^2(t) = \sigma_0^2+\Xi(t)$. The proof follows via the application of Eq.~(\ref{eq:W}): with a time-dependent $\sigma^2(t)$, only the second cumulant is time dependent,
\be
\mu_2(t) = \kappa_2(t)  = \mu_0^2+\sigma^2(t) =  \mu_0^2+\sigma_0^2 +\Xi(t),
\ee
while all the higher-order cumulants are constant. 
The drift of the backward process, i.e.\ the score, then follows from
\be
-\partial_x\log P(x,t) = \frac{x}{\sigma^2(t)} -\frac{\mu_0}{\sigma^2(t)} \tanh\left(\frac{\mu_0 x}{\sigma^2(t)} \right),
\ee
with $t\to T-\tau$.
Note that this implies that all the terms in a polynomial expansion of the drift depend on time via $\sigma^2(t)$, but that these are resummed in such a way that only the second cumulant is time dependent. 

\begin{figure}[t]
\begin{center}
\includegraphics[width=0.32\textwidth]{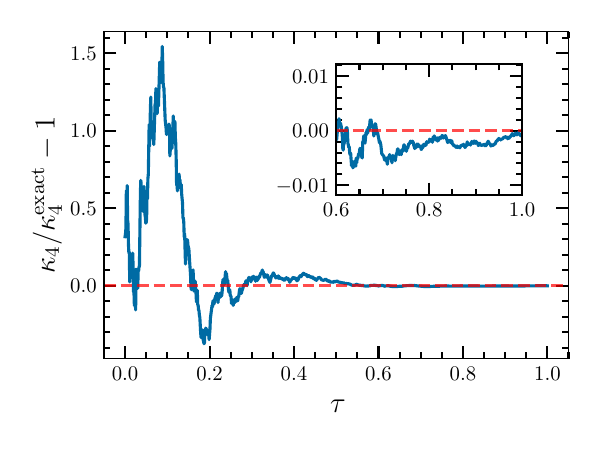} 
\includegraphics[width=0.32\textwidth]{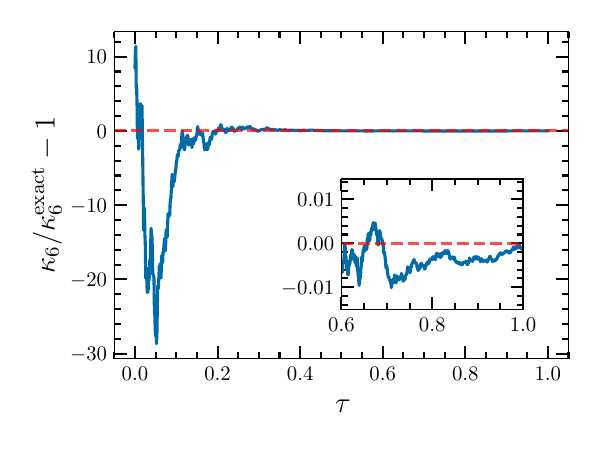}
\includegraphics[width=0.32\textwidth]{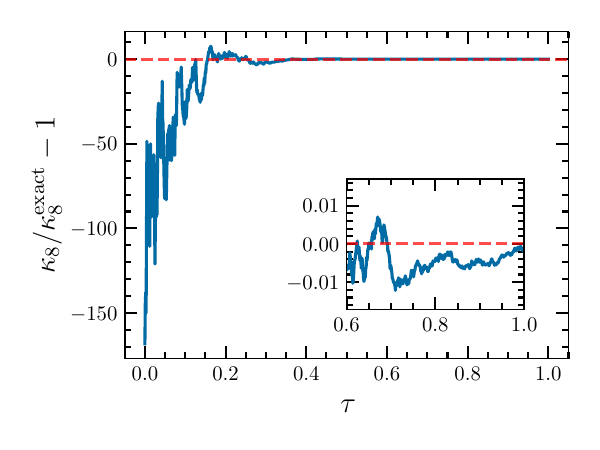} 
\end{center}
  \caption{As in the preceding figure, 
  employing the analytical score during the backward process, using $10^6$  trajectories. 
  }
  \label{fig:2peak-bw-cum}
\end{figure}

In Fig.~\ref{fig:2peak-bw-cum}, we show the evolution of the normalised cumulants during the backward process, using the analytical score with $10^6$  trajectories -- the second moment is shown in Fig.~\ref{fig:2peak-2nd-mom} (right). We observe the same behaviour as in the case with the learned score, see Fig.~\ref{fig:2peak-bw-cum-DM} (below).
We hence conclude that the observed fluctuations during the first part of the backward process are due to the finite number of trajectories and not due to a poorly learned score, which is reassuring.

To assess quantitatively how well the cumulants are determined, we have estimated the first four nonvanishing cumulants $\kappa_n$, with the statistical uncertainty determined by a bootstrap analysis. The results are shown in Table~\ref{table:two-peak}. The first row contains the exact values and the second row the values generated by direct sampling (target ensemble). The third row contains the results obtained with the diffusion model in the variance-expanding scheme, as discussed above. We observe excellent agreement for the higher-order cumulants. For the second cumulant a small deviation is observed, which is however less than 1\%. It is interesting that better agreement is observed for the higher-order cumulants.

\begin{table}[h]
\centering
\begin{tabular}{l|c|c|c|c}
\hline\hline
                                & $\kappa_2$    & $\kappa_4$    & $\kappa_6$    & $\kappa_8$     \\ \hline
Exact                           & $1.0625$      & $-2$          & $16$          & $-272$           \\ 
Data                            & $1.0624(5)$   & $-2.000(2)$   & $16.00(2)$  & $-272.0(6)$   \\          
Variance expanding              & $1.0692(6)$   & $-2.001(2)$   & $16.03(3)$  & $-272.7(6)$    \\
Variance preserving (DDPM)      & $1.0609(5)$   & $-1.976(2)$   & $15.72(2)$  & $-265.6(6)$    \\ 
%
\hline\hline
\end{tabular}

\caption{First four nonvanishing cumulants $\kappa_n$ in the two-peak model, as obtained from training data and from diffusion models without a drift (variance expanding) and with a drift (variance preserving, DDPM).
Statistical errors are computed by bootstrap resampling of a $10^6$ size dataset with 1000 bins.
}
\label{table:two-peak}
\end{table}

We conclude that the higher-order cumulants are learned correctly, within numerical uncertainty, with a noticeable effect for a finite number of trajectories near the start of the backward process. One may infer from this behaviour that the first part of the backward process is not that relevant for the evolution towards the final stages, which is worth exploring further.

\subsection{Denoising diffusion probabilistic models}

Next we turn to the class known as denoising diffusion probabilistic models, or DDPMs, in the continuous-time limit. These models have a nonzero drift, which leads to qualitatively different time dependence: the distribution at the end of the forward process is expected to be a normal distribution, and hence all the higher-order cumulants should go to zero. 
We use a linear drift, with coefficient $k(t) = g^2(t)$. The qualitative features do not depend on the choice for $g(t)$, but for the numerics shown below we have taken $g(t) = \sigma^{t/T}$, with $\sigma=10, T=1$.
The specific choice of $g(t)$ determines the time profile, via the definition
\be
u(t) = \int_0^tds\, g^2(s) = \frac{T}{\log \sigma^2}\left[ \sigma^{2t/T}-1\right].
\ee
Also in this case the analytical score is available and the time-dependent distribution reads
\be
\label{eq:pxt2}
P(x,t) = \half \left[ \cN(x; \mu(t),\sigma^2(t)) + \cN(x; -\mu(t),\sigma^2(t)) \right],
\ee
where
\be
\mu(t)=\mu_0 f(t,0),
\qqquad
\sigma^2(t) = \sigma_0^2 f^2(t,0) + \Xi(t),
\ee
with
\begin{align}
f(t,s) &= e^{-\half\int_s^tds'\, k(s')} = e^{-\half u(t)+\half u(s)}, 
\\
\Xi(t) &= \int_0^tds\, f^2(t,s)g^2(s) = 1-f^2(t,0).
\end{align}
The solution (\ref{eq:pxt2}) describes the evolution of the distribution during the forward process. The proof is the same as above; with this distribution the second and higher cumulants evolve as
\begin{align}
\label{eq:kapp2DDPM}
\kappa_2(t) &= \mu^2(t) + \sigma^2(t) = \left(\mu_0^2 + \sigma_0^2 -1\right) f^2(t,0) + 1, \\
\kappa_{n>2}(t) &= \kappa_n(0) f^n(t,0),
\end{align}
as it should be.
At $\tau=T$, $\kappa_2(T)\to 1$ and $\kappa_{n>2}(T)\to 0$, up to exponentially suppressed terms. The distribution then becomes normal, $P(x,T) = \cN(x;0,1)$, again up to exponentially suppressed terms, see also App.~\ref{app:Gaussian}.

\begin{figure}[h]
\begin{center}
\includegraphics[width=0.4\textwidth]{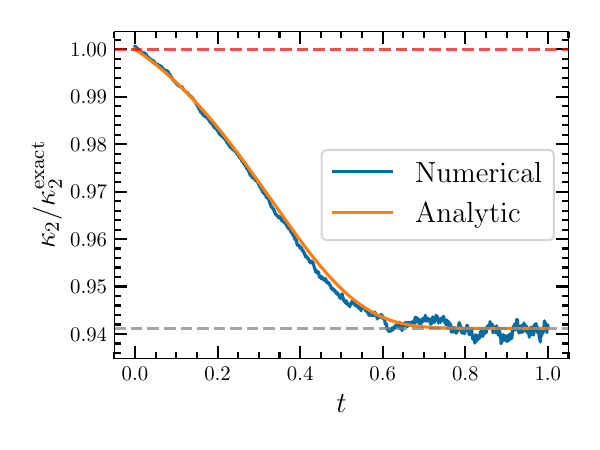} 
\includegraphics[width=0.4\textwidth]{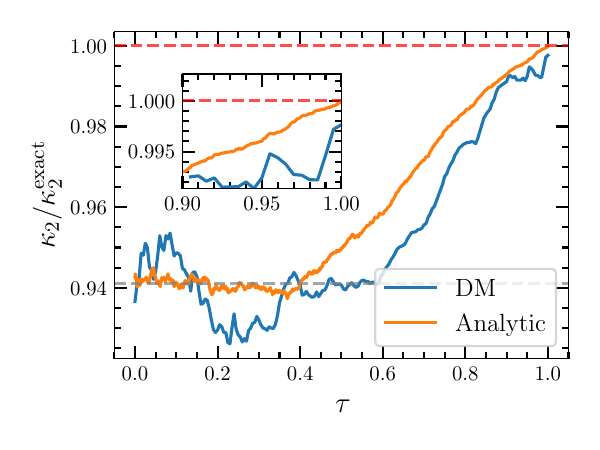}
\end{center}
  \caption{Evolution of the normalised $2^{\rm nd}$ cumulant,
 presented as $\kappa_2/\kappa_2^{\rm exact}$,
  in the two-peak model in the DDPM, during the forward process (left) and during the backward process (right), with the score determined by the diffusion model and with the analytical score, using $10^6$ trajectories in all cases. In the forward process, the analytical solution is shown as well. Other parameters as above. 
  }
  \label{fig:2peak-bw-cum2-DM-ddpm}
\end{figure}

The evolution of the second cumulant is shown in Fig.~\ref{fig:2peak-bw-cum2-DM-ddpm}. 
Note that the cumulant is normalised with the target value, $\kappa_2(0)=\mu_0^2+\sigma_0^2 = 1.0625$, such that the expected value at the end of the forward process is $1/1.0625=0.9412$. In the forward process we also show the analytical solution (\ref{eq:kapp2DDPM}), to make clear that the observed noisy behaviour is due to the finite number of trajectories. For the backward process, we show the evolution using the learned score and the analytical score from the time-dependent solution (\ref{eq:pxt2}), using an equal number of trajectories. The behaviours are initially somewhat different, but for $\tau/T>0.5$ both processes converge towards the target value.

\begin{figure}[!htbp]
\begin{center}
\includegraphics[width=0.32\textwidth]{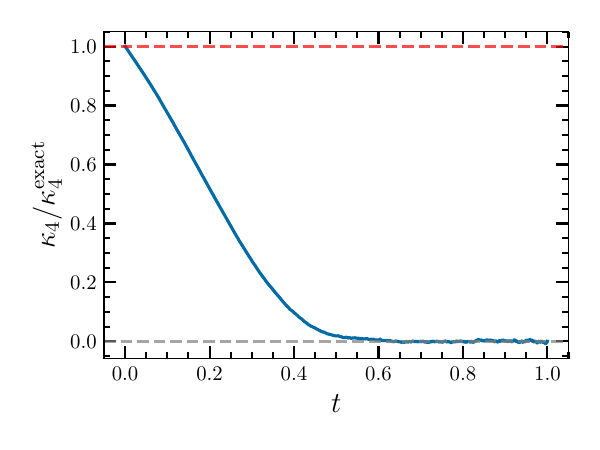} 
\includegraphics[width=0.32\textwidth]{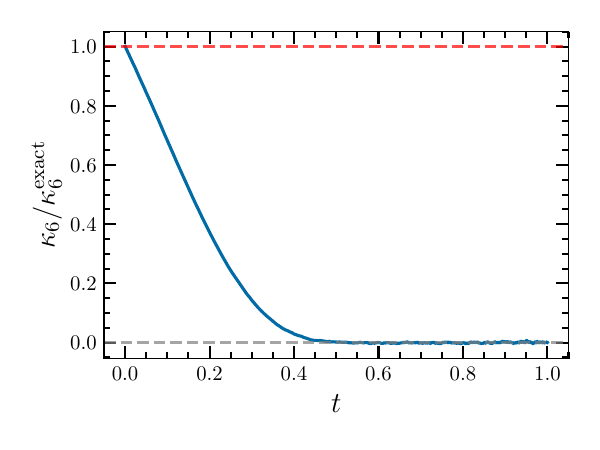}
\includegraphics[width=0.32\textwidth]{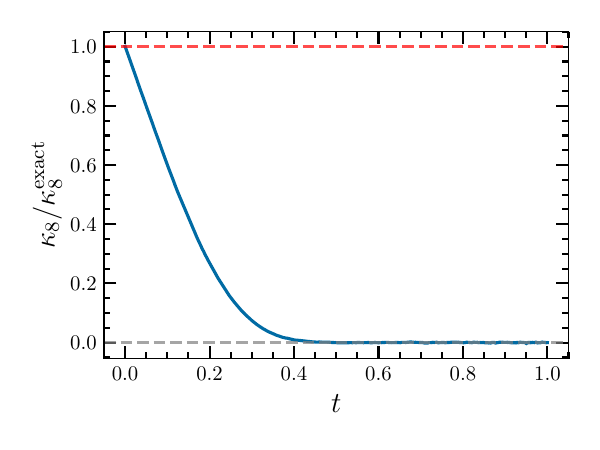}
\end{center}

\begin{center}
\includegraphics[width=0.32\textwidth]{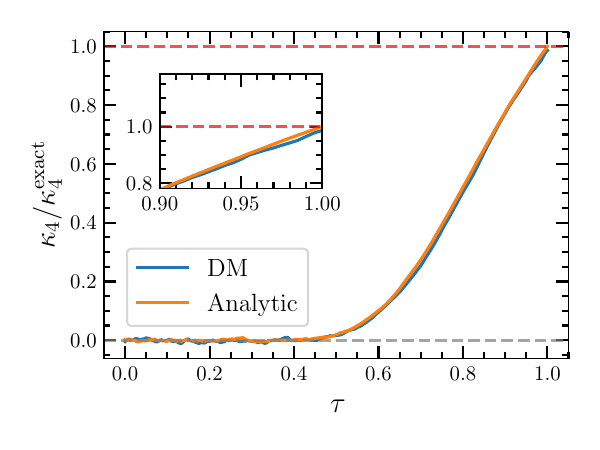} 
\includegraphics[width=0.32\textwidth]{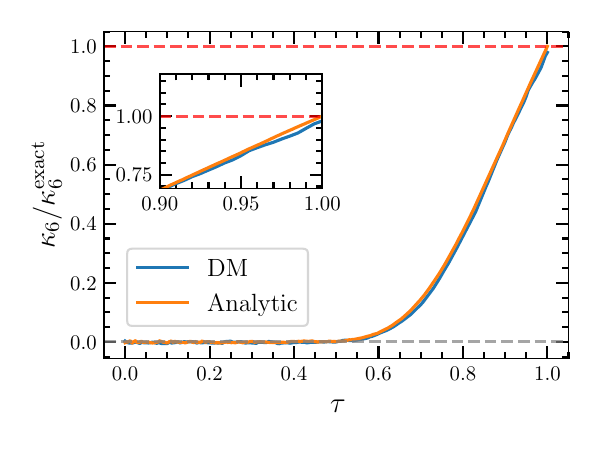}
\includegraphics[width=0.32\textwidth]{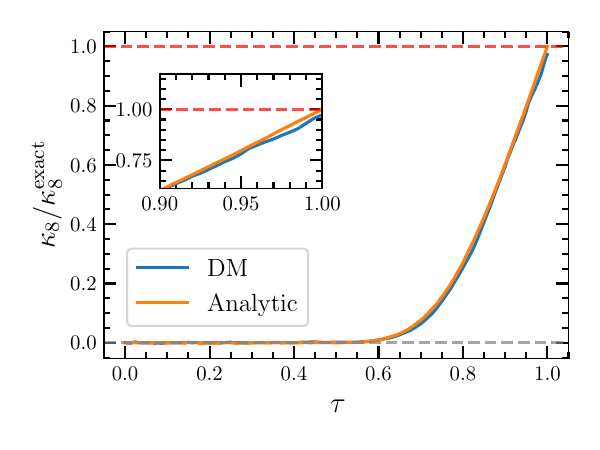} 
\end{center}
  \caption{Evolution of the normalised $4^{\rm th}, 6^{\rm th}$ and $8^{\rm th}$ cumulants,
  presented as $\kappa_n/\kappa_n^{\rm exact}$,
  in the two-peak model in the DDPM, during the forward process (above), and during the backward process (below), with the score determined by the diffusion model and with the analytical score, using $10^6$ trajectories in all cases. Other parameters as above. 
  }
  \label{fig:2peak-bw-cum-DM-ddpm}
\end{figure}

Higher-order cumulants are presented in Fig.~\ref{fig:2peak-bw-cum-DM-ddpm}, for the forward (top row) and backward (bottom row) process. The cumulants evolve from the target value to zero, and vice versa, in a much more controlled manner than in the variance-expanding scheme; the cancellations required above are not needed here.
The results, with statistical error, are given in Table~\ref{table:two-peak}, fourth row. We note an agreement which is slightly worse compared to the variance-expanding scheme. This may be due to the qualitatively different behaviour observed during the evolution: while fluctuations are suppressed, there is a stronger dependence on the value obtained around, say, $\tau/T=0.5$, to reach the expected value at $\tau=T$.  
It would be interesting to combine features of the two schemes, as the effects due to a finite number of trajectories affect the second cumulant and higher cumulants in opposite ways.

\section{Lattice field theory}
\label{sec:lattice}

We now move to the case of many degrees of freedom and consider a Euclidean field theory, discretised on a lattice. For a real scalar field $\phi(x)$, the target probability distribution reads
\be
P_0[\phi] = \frac{1}{Z} e^{-S[\phi]}, \qqquad
Z = \int D\phi\, e^{-S[\phi]},
\ee
where $S[\phi]$ is the Euclidean action and $Z$ denotes the partition function. The integration is over all field configurations and $D\phi$ denotes the discretised measure,
\be
\int D\phi = \prod_x \int_{-\infty}^\infty d\phi_x,
\ee
where the product is over all spacetime points, forming a hypercubic (square) lattice in $n$ (two) dimensions. The simplest interacting field theory is a $\lambda\phi^4$ theory, with continuum action
\be
\label{eq:Scon}
S[\phi] = \int d^nx\, \left[ \half \sum_{\nu=1}^n\left(\partial_\nu\phi(x)\right )^2 +\half m_0^2\phi^2(x) +\frac{1}{4}\lambda_0\phi^4(x)\right].
\ee
As stated we use Euclidean signature, with $\partial_\nu$ ($\nu=1,\ldots,n$) denoting partial derivatives. Following the standard route to discretisation \cite{Smit:2002ug}, the corresponding lattice action reads
\be
\label{eq:Slat}
S[\phi] = \sum_x \left[ -2\kappa\sum_{\nu=1}^n \phi_x \phi_{x+\hat \nu} + \phi_x^2 + \lambda\left(\phi_x^2 -1\right)^2 \right].
\ee
We use periodic boundary conditions. 
The relation between the continuum and lattice fields, the lattice spacing, and the parameters $m_0, \lambda_0$ and $\kappa, \lambda$ can be found in Refs.\ \cite{Smit:2002ug,Wang:2023exq}. Note that $\kappa$ is the so-called hopping parameter, not to be confused with a cumulant. The actions (\ref{eq:Scon}, \ref{eq:Slat}) are non-Gaussian and hence lead to nonvanishing higher-order cumulants. Below we denote the field with $\phi(x)$.

Returning to the diffusion model, the equations discussed above for one degree of freedom can be taken over directly. The forward process reads
\be
\partial_t \phi(x,t)  = K[\phi(x,t),t] + g(t)\eta(x,t), 
\ee
with the backward process 
\be
\partial_\tau\phi(x,\tau) = -K[\phi(x,\tau),T-\tau] + g^2(T-\tau)\nabla_\phi \log P(\phi,T-\tau) + g(T-\tau)\eta(x,\tau).
\ee
Here $K[\phi,t]$ is the possible drift term and $\eta\sim \cN(0,1)$ is Gaussian noise with variance 1, applied locally at each lattice coordinate, i.e.\ 
\be
\label{eq:noise}
\E_\eta[ \eta(x,s)  \eta(x',s')]=\delta(s-s')\delta(x-x').
\ee
Note that $g(t)$ still only depends on time, but one could introduce $x$ dependence as well. As above, we assume the first moment vanishes, or has been subtracted, $\phi(x)\to \phi(x)-\E_{P_0}[\phi(x)]$.

With a linear drift, $K[\phi(x,t),t] =- \half k(t)\phi(x,t)$, and the initial condition $\phi_0\sim P_0[\phi_0]$, the forward equation is solved by
\be
\phi(x,t)  = \phi_0(x) f(t,0) + \int_0^t ds\, f(t,s)g(s) \eta(x,s),
\ee
where $f(t,s)$ was defined in Eq.\ (\ref{eq:fts}).
The equal-time two-point function (or two-point correlation function or propagator) then reads
\be
\label{eq:Gt}
G(x,y;t)\equiv \E[\phi(x,t)\phi(y,t)] = \E_{P_0} [ \phi_0(x)\phi_0(y) ] f^2(t,0) + \Xi(t)\delta(x-y),
\ee
where 
\be
G_{\rm target}(x,y) \equiv  \E_{P_0}[\phi_0(x)\phi_0(y)]
\ee
is the full two-point function in the target theory. $\Xi(t)$ is exactly the same as in Eq.~(\ref{eq:Xi}), having used Eq.~(\ref{eq:noise}).

Here we focus on moments and cumulants, involving products of the field at coinciding spacetime points -- we come back to the propagator (\ref{eq:Gt}) below.  Moments are then defined as 
\be
\mu_n(x,t) = \E[\phi^n(x,t)].
\ee
Under the usual assumption that the target theory is translationally invariant, moments and cumulants are independent of $x$ and the $x$-label may be dropped.

Since the noise is applied at each spacetime point separately, the computation for the moments and cumulants is exactly as before. We hence give immediately the results for the generation functionals. Moments are generated by 
\be
Z[J] = \E[e^{J(x,t)\phi(x,t)}] = e^{\half J^2(x,t) \Xi(t)} \int D\phi_0\, P_0[\phi_0] e^{J(x,t) \phi_0(x)f(t,0)},
\ee 
and the cumulant-generating function reads
\be
W[J] = \log Z[J] = \half J^2(x,t) \Xi(t) + \log \int D\phi_0\, P_0[\phi_0] e^{J(x,t) \phi_0(x)f(t,0)}.
\ee 
The second moment or cumulant is given by Eq.\ (\ref{eq:Gt}), evaluated at $x=y$.\footnote{The delta function should be understood as defined on the discretised lattice, $\delta(x-y)\to \delta_{x,y}$, where $\delta_{x,y}$ is the Kronecker delta with $ \delta_{x,x}=1$.}
All higher-order cumulants are given by
\be
\kappa_{n>2}(t) = \frac{\delta^n W[J]}{\delta J(x,t)^n} \Big|_{J=0} =  \frac{\delta^n}{\delta J(x,t)^n} \log \E_{P_0}[ e^{J(x,t)\phi_0(x)f(t,0)}]\Big|_{J=0},
\ee
and are hence equal to the cumulants in the target theory, multiplied with the time-dependent function $f^n(t,0)$. In particular, for pure diffusion we find again that 
\be
\kappa_{n>2}(t) = \kappa_n(0) \qqquad (\mbox{pure diffusion}).
\ee

\begin{figure}[t]
\begin{center}
\includegraphics[width=0.48\textwidth]{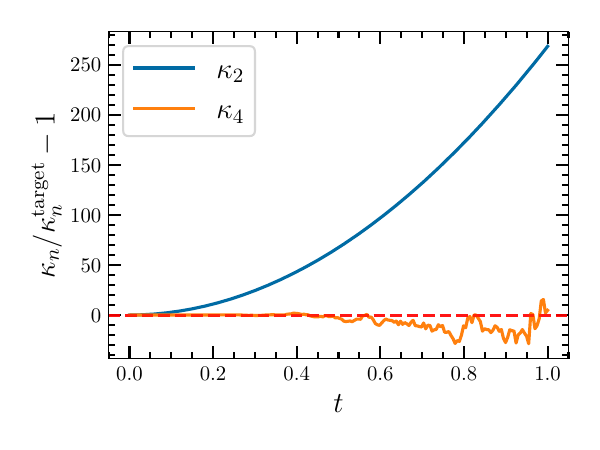} 
\includegraphics[width=0.48\textwidth]{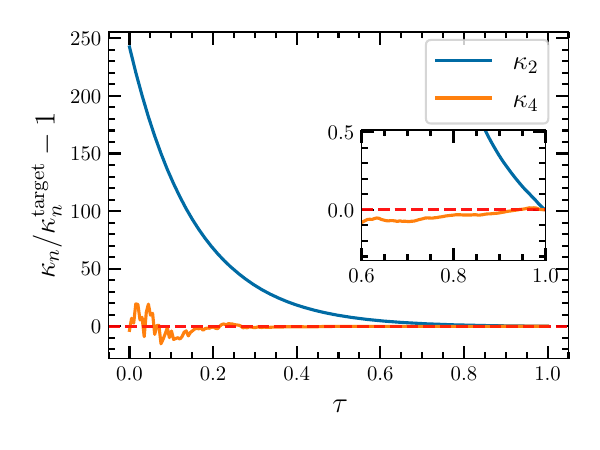}
\end{center}
  \caption{Evolution of the normalised $2^{\rm nd}$ and $4^{\rm th}$ cumulant,
  presented as $\kappa_n/\kappa_n^{\rm target}-1$,
  in the two-dimensional $\phi^4$ theory, during the forward (left) and backward (right) process with the score determined by the diffusion model. 
  }
  \label{fig:phi4-fw-bw-cum-DM}
\end{figure}

\begin{figure}[t]
\begin{center}
\includegraphics[width=0.48\textwidth]{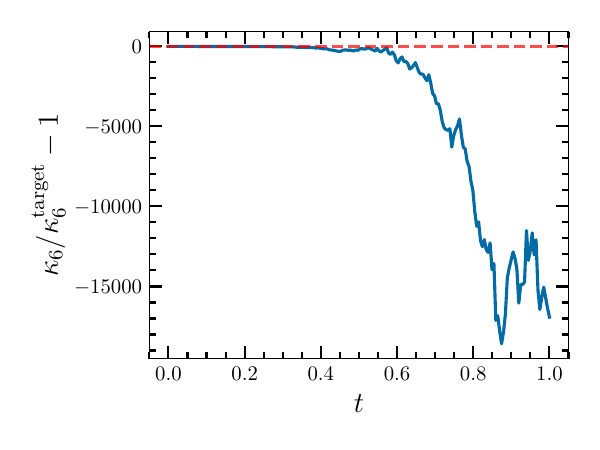} 
\includegraphics[width=0.48\textwidth]{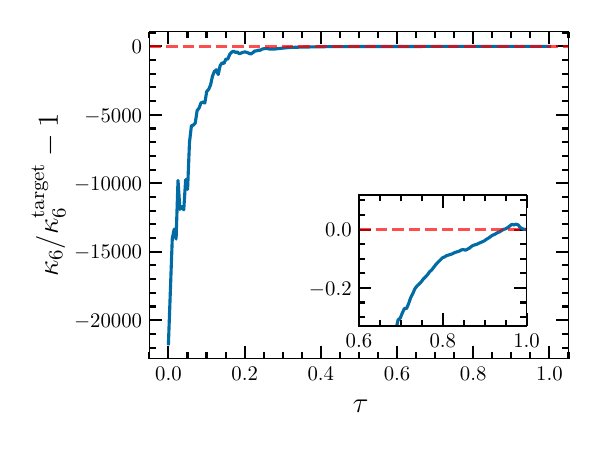}
\end{center}
  \caption{Evolution of the normalised $6^{\rm th}$ cumulant,
  presented as $\kappa_6/\kappa_6^{\rm target}-1$,
  in the two-dimensional $\phi^4$ theory, during the forward (left) and backward (right) process with the score determined by the diffusion model. 
  }
  \label{fig:phi4-fw-bw-cum-DM-6}
\end{figure}

We will now verify these results in the scalar field theory introduced above, defined on a two-dimensional lattice, using the implementation of the variance-expanding diffusion model without drift, previously discussed in this context in Refs.\ \cite{Wang:2023exq,Wang:2023sry}.
The results shown below are obtained on a $32\times 32$ lattice, with hopping parameter $\kappa=0.4$  and coupling $\lambda=0.022$. The theory is in the symmetric phase. We used $10^5$ configurations to train the model using the variance-expanding scheme with $\sigma=25$ and $T=1$, and also $10^5$ configurations to evolve the cumulants during the forward and backward processes.

Fig.\ \ref{fig:phi4-fw-bw-cum-DM} shows the $2^{\rm nd}$ and $4^{\rm th}$ cumulant, normalised with the numerically computed target value, during the forward (left) and backward (right) process. As expected, the $2^{\rm nd}$ moment or cumulant increases (decreases) as in Eq.~(\ref{eq:kappa2exact}). During the backward process, the second cumulant  approaches the target result linearly from above, as expected, see App.~\ref{app:Gaussian}. 
The $4^{\rm th}$ cumulant is approximately constant, with the effect of a finite number of trajectories visible towards the end (start) of the forward (backward) process, as above. 
The forward and backward evolution of the $6^{\rm th}$ order cumulant is shown in Fig.~\ref{fig:phi4-fw-bw-cum-DM-6}. The cumulant is approximately conserved during the first (second) part of the forward (backward) evolution and suffers from incomplete cancellations for the remainder, similar to the evolution in the two-peak model. 
We note that with $32^2\sim 10^3$ lattice sites, the total number of trajectories is $32^2\times 10^5\sim 10^8$.
The insets show the evolution at the end of the backward process.

\begin{table}[h!]
\centering
\begin{tabular}{l|c|c|c|c}
\hline\hline
            & $\kappa_2$     & $\kappa_4$       & $\kappa_6$       & $\kappa_8$     \\ \hline
HMC (normalised)    & $0.39597(4)$   & $-0.29453(6)$     & $0.90108(28)$   & $-5.8689(25)$    \\ 
Diffusion model     & $0.39598(4)$   & $-0.29454(7)$     & $0.90113(32)$   & $-5.8694(28)$    \\ 
%
\hline\hline
\end{tabular}

\caption{First four nonvanishing cumulants $\kappa_n$ in the scalar $\phi^4$ field theory, with $\kappa$ = 0.4, $\lambda$ = 0.022 and $10^5$ configurations on a $32^2$ lattice, using normalised HMC data and as obtained from the diffusion model in the variance-expanding scheme.
 Statistical errors are computed by bootstrapping.
 }
 \label{table:phi4}
\end{table}

Estimates for the first four nonvanishing cumulants are presented in Table~\ref{table:phi4}, with statistical uncertainty. We note here that the cumulant values are given for the normalised data. To revert to the unnormalised (i.e.\ original) data, the $n^{\rm th}$ cumulant $\kappa_n$ has to be multiplied with the $n^{\rm th}$ power of $\phi_{\rm max}=-\phi_{\rm min}$,  see App.~\ref{app:DM} for details. 
Importantly, we observe that the cumulants are reproduced to high precision, including $\kappa_2$.
This illustrates that the diffusion model is capable of learning higher-order cumulants for a system with many degrees of freedom, such as a lattice field theory.

Returning finally to the propagator (\ref{eq:Gt}) in the case without a drift,  $f(t,0)=1$,  and using translational invariance,  we note that it reads in momentum space 
\be
G(p,t) = G_{\rm target}(p) +\Xi(t) = \frac{1}{p^2+m^2 + \Sigma(p)} +\Xi(t).
\ee
Here we used the Dyson equation for the full propagator of the target theory, 
\be
G_{\rm target}^{-1}(p) = G_0^{-1}(p) +\Sigma(p), \qqquad G_0^{-1}(p) = p^2+m^2.
\ee
We note that during the forward process, with $\Xi(t)$ increasing, the most ultraviolet (large momentum) modes are destroyed first, and reversely, during the backward process,  with $\Xi(t)$ decreasing, the most infrared (small momentum) modes are denoised first. The direction of this flow of information (from large spatial scales to small ones, or vice versa) is interesting to explore further, e.g.\ in the context of the Renormalisation Group (see e.g.\ Refs.~\cite {cotler2023renormalizingdiffusionmodels, Berman:2022uov, Berman:2023rqb} for connections between diffusion and (inverse) Renormalisation Group flows) or of nonequilibrium phase transitions and symmetry breaking/restoration (see e.g.\ Ref.~\cite{sclocchi2024phasetransitiondiffusionmodels} for phase transitions in diffusion models applied to ImageNet data).

\section{Conclusion and outlook}
\label{sec:con}

In this paper we investigated how higher-order moments and cumulants are learnt in diffusion models, by deriving exact expressions for the moment- and cumulant-generating functionals. We have demonstrated that higher-order cumulants are exactly conserved in models without a drift, such as the variance-expanding scheme. The distribution at the end of the forward process is therefore as correlated as the target distribution. In models with a drift, such as DDPMs, higher-order cumulants go to zero and the distribution at the end of the forward process is normal, up to exponentially suppressed terms. In both cases, the score incorporates the knowledge of the higher-order correlations, which are therefore regenerated during the backward process, when starting from a normal distribution. 
These predictions were subsequently verified in an exactly-solvable but nontrival model with one degree of freedom and in a lattice scalar field theory. The use of the latter is highly relevant for the application of diffusion models to generate field configurations as an alternative to standard Monte Carlo-based approaches. Since higher-order cumulants contain the information on interactions between fundamental degrees of freedom, it is indeed of utmost importance that these can be encoded and learnt properly. 

In the numerical implementation of the variance-expanding scheme, we observed that at the final (initial) stages of the forward (backward) process the higher-order cumulants suffer from incomplete cancellations between trajectories and therefore appear ``noisy''. We have demonstrated that this is not due to a poorly learned score but due to finite statistics. It will therefore be interesting to investigate how relevant the first stage of the backward process is in the generation of new configurations and whether efficiency can be gained by starting the backward process slightly later. For efficiency gains by adapting the noise scheduler, see e.g.\ 
Refs.~\cite{chen2023importancenoiseschedulingdiffusion,hang2024improvednoiseschedulediffusion,ikeda2024speedaccuracytradeoffdiffusionmodels}.
We also note here that we have not incorporated any acceptance/reject step in this study, which is possible in principle \cite{Wang:2023exq} and a further area to explore in cases where minor deviations between the target and generated data need to be addressed.
Finally, an analysis of the two-point function in field theory indicated an interesting interplay between momentum scales in the propagator and the scale of the noise during the forward and backward processes, which is worth exploring further, e.g.\ using the framework of the Renormalisation Group or nonequilibrium phase transitions.

Although the main thrust of our paper is theoretical, looking forward more generally we note that understanding the preservation and learning of complex correlations in diffusion models, as encoded in higher-order cumulants, might inform the development of more robust generative models across various domains. Additionally, modifying the training or sampling procedures of diffusion models to explicitly account for these may enhance their performance in capturing intricate data structures. Such advancements could lead to more accurate and efficient models in fields ranging from image and signal processing to the simulation of complex physical systems, including lattice field theory.

\noindent
{\bf Acknowledgements} --  
GA is supported by STFC Consolidated Grant ST/X000648/1.
DEH is supported by the UKRI AIMLAC CDT EP/S023992/1.
LW thanks the DEEP-IN working group at RIKEN-iTHEMS for its support in the preparation of this paper.
KZ is supported by the CUHK-Shenzhen university development fund under grant No.\ UDF01003041 and UDF03003041, and Shenzhen Peacock fund under No.\ 2023TC0179.

We acknowledge the support of the Supercomputing Wales project, which is part-funded by the European Regional Development Fund (ERDF) via Welsh Government.

\noindent
{\bf Research Data and Code Access} --
The code and data used for this manuscript are available from Ref.~\cite{habibi_2024_14041604}.

\noindent
{\bf Open Access Statement} -- For the purpose of open access, the authors have applied a Creative Commons Attribution (CC BY) licence to any Author Accepted Manuscript version arising.

\appendix
\renewcommand{\theequation}{\Alph{section}.\arabic{equation}}

\section{Implementation of the diffusion model}
\label{app:DM}

In this Appendix, we give some details on the numerical implementation of the diffusion model. 

\textbf{Estimating the score --} To estimate the score of a distribution we may train a score-based model based on some sample dataset of the target distribution with score-matching \cite{Hyvarinen:2005jmlr}.
The score function approximation $s_\theta(x,t)$ can be trained with the Fisher divergence objective such as in Ref.\ 
\cite{Song:2021scorebased},
 \be
 \mathcal{L}(\theta, \lambda) := \frac{1}{2}\int_0^T dt\, \E_{P_t(x)}\left[\lambda(t) \bignorm s_\theta(x,t) - \nabla\log P_t(x)\bignorm^2_2\right],
  \label{eq: weighted_training_objectice}
 \ee
where the weight $\lambda(t)$ is chosen to be the variance of the noise at time $t$ which, for $g(t) = \sigma^{t/T}$, equals
 \be
 \lambda(t) = \frac{T}{2\log\sigma}\left(\sigma^{2t/T} - 1\right).
 \ee
A practical and computationally easier training objective is to approximate the score function of the marginal probability at some time $t$, $\nabla\log P_t(x)$, with the score of a transition kernel $\nabla\log P_t(x_t | x_0)$. In the case of an affine drift $K(x, t)$, the transition kernel is always a Gaussian distribution kernel such that $ P_t(x_t | x_0) = \mathcal{N}(x_t; x_0, \lambda(t))$, and $P_t(x_t) = \int dx_0\, P_t(x_t | x_0)P_\text{data}(x_0)$.

In a given epoch, we sample a random time step from the uniform distribution, $t \sim \mathcal{U}(\varepsilon, T)$, with $\varepsilon$ close to $0$, for every batch in the data set and perturb our sample $x_0$ with noise $\mathcal{N}(0, \lambda(t))$. For an affine drift, the score of the transition kernel in the training objective can be written as $\nabla\log P_t(x_t|x_0) = -(x_t - x_0)/\lambda(t)$, and using Eq.~(\ref{eq: weighted_training_objectice}), we obtain the simplified loss function
 \begin{align}
 \nn
      \mathcal{L}(\theta, \lambda) &= \frac{1}{2}\sum_{t=0}^T \, \E_{P_t(x)}\left[\lambda(t) \Bignorm s_\theta(x,t) + \frac{x_t - x_0}{\lambda(t)}\Bignorm^2_2\right] \\
      &= \frac{1}{2}\sum_{t=0}^T \, \E_{P_t(x)}\left[\bignorm s_\theta(x,t) \sqrt{\lambda(t)} + z_t \bignorm^2_2\right], \qqquad z_t\sim\mathcal{N}(0,1).
 \end{align}
Having a trained score model $s_\theta^*(x,t)$, we can obtain samples from the target distribution by numerically solving the backwards stochastic process, c.f.\ Eq.~\ref{eq:bw},
\be
x_{\tau + \Delta \tau} = x_\tau +\left[ - K(x_\tau, T-\tau) + g^2(T-\tau)s_\theta^*(x_\tau, T-\tau) \right] \Delta \tau + g(T-\tau)\sqrt{\Delta \tau}\ \eta_\tau,
\ee
where $\Delta\tau$ is the step size, $\eta_\tau \sim \mathcal{N}(0,1)$ and the equation is solved from $\tau = 0$ to $T$.

\textbf{Two-peak model with one degree of freedom -- } 
To model the score $s_\theta(x,t)$, we use a fully connected neural network conditioned on the time information using Gaussian Fourier feature mapping \cite{tancik2020:GaussianFourier}. For inference, we choose to run the backward process using 1000 steps for $10^6$ trajectories.
Our choice of hyperparameters is summarised in table \ref{tab:model_training_hyperparams}.

\begin{table}[h!]
\centering
\begin{tabular}{ll|ll}
\toprule
\multicolumn{2}{c}{\textbf{Model Hyperparameters}} & \multicolumn{2}{c}{\textbf{Training Hyperparameters}} \\ \midrule
\textbf{Hyperparameter}    & \textbf{Value}         & \textbf{Hyperparameter}    & \textbf{Value}         \\ \midrule
Layers                     & [64, 64]               & Learning Rate              & 1e-4                   \\
Time Embedding dims        & 128                    & Batch Size                 & 512                    \\
Activation Function        & LeakyReLU              & Optimizer                  & Adam                   \\
Weight Initialization      & PyTorch default        & Max Epochs                 & 200                    \\ \bottomrule \\
\end{tabular}

\caption{Model and training hyperparameters used in the training of the two-peak model. We save the weights with the best loss during the training process and set the training to stop early if the loss has not improved within 50 epochs. An early stop was observed occurring after an average of 100 epochs for a dataset of $10^6$ realisations.}
\label{tab:model_training_hyperparams}
\end{table}

 \textbf{Lattice field theory -- } Here we use the same setup as in Ref.~\cite{Wang:2023exq}, see App.~A of that paper.

For training purposes, the HMC data has been normalised using the reversible transformation
\be
\tilde\phi(x) = 2\left(\frac{\phi(x)-\phi_{\rm min}}{\phi_{\rm max} - \phi_{\rm min}} - \half\right),
\qquad
\phi(x) = \half\left(\tilde\phi(x) +1\right)\left( \phi_{\rm max} - \phi_{\rm min} \right) + \phi_{\rm min},
\ee
where $\phi_{\rm min, max}$ are the minimal and maximal value of the field over the entire ensemble for fixed lattice parameters. For a symmetric distribution, $n^{\rm th}$ order moments and cumulants for unnormalised and normalised data are related via a multiplication or division by $\phi_{\rm max} = -\phi_{\rm min}$. For our ensemble $\phi_{\rm max} \sim - \phi_{\rm min} = ~ 5.711$.

\section{Gaussian target distribution}
\label{app:Gaussian}

In this Appendix we discuss the case of a Gaussian target distribution in some detail. Since the distribution is a Gaussian throughout the forward and backward process (with vanishing mean), the only dynamical quantity is the variance, i.e.\ the second moment or cumulant. 
It is important to distinguish the variances during the process:
\begin{itemize}
\item  variance of the target distribution: $\sigma^2_{\rm target}$,
\item  initial condition of the forward process: $\sigma^2_{\rm fw}(0)=\sigma^2_{\rm target}$,
\item (final) variance during the forward process: $\sigma^2_{\rm fw}(t), \sigma^2_{\rm fw}(T)$,
\item initial condition of the backward process: $\sigma^2_{\rm bw}(0) = \sigma^2_{0, \rm bw}$,
\item (final) variance during the backward process: $\sigma^2_{\rm bw}(\tau), \sigma^2_{\rm bw}(T)$.
\end{itemize}
A diffusion model works well if the final result of the backward process is (approximately) equal to the variance of the target distribution, $\sigma^2_{\rm bw}(T)\sim \sigma^2_{\rm target}$.

We consider the forward process
\be
\dot x(t) = -\frac{1}{2}k(t)x(t)  + g(t)\eta(t),
\ee
where the drift includes the cases considered above: $k(t)=0, g^2(t)$. The corresponding Fokker-Planck equation (FPE) reads
\be
\partial_t P(x,t) =  \frac{1}{2}\partial_x\left[ g^2(t) \partial_x  + k(t) x \right] P(x,t).
\ee
With a linear drift, the solution is a Gaussian distribution,
\be
P(x,t) = \frac{e^{-x^2/2\sigma^2_{\rm fw}(t)}}{\sqrt{2\pi \sigma^2_{\rm fw}(t)}},
\qqquad
\sigma^2_{\rm fw}(t) = \E_{P(x,t)}[ x^2(t) ]  = \int dx\, P(x,t) x^2.
\ee
Substituting this Ansatz in the FPE yields the equation $\sigma^2_{\rm fw}(t)$ has to satisfy,
\be
\label{eq:sigmadot}
\dot \sigma^2_{\rm fw}(t) = - k(t)\sigma^2_{\rm fw}(t) + g^2(t).
\ee
Using the notation in Sec.\ \ref{sec:mom}, the equation above is solved by
\be
\label{eq:sigma2}
\sigma^2_{\rm fw}(t) = \sigma^2_{\rm target} f^2(t,0) + \Xi(t),
\ee
in terms of the initial variance $\sigma^2_{\rm target}$, and 
\be
f(t,s) = e^{-\half\int_s^t ds'\, k(s')},
\qqquad
\Xi(t) =  \int_0^tds\, f^2(t,s)g^2(s).
\ee
The backward process (with $\tau=T-t$) is
\be
 x'(\tau) = \frac{1}{2}k(T-\tau)x(\tau) +g^2(T-\tau) \partial_x\log P(x,T-\tau)  + g(T-\tau)\eta(\tau).
\ee
With
\be
\partial_x\log P(x,t) = -x(t)/\sigma^2_{\rm fw}(t),
\ee
the drift for the backward process is linear, such that
\be
 x'(\tau) = -\half k_{\rm bw}(\tau) x(\tau)   + g(T-\tau)\eta(\tau),
\ee
with the coefficient
\be
k_{\rm bw}(\tau) =   \frac{2g^2(T-\tau)}{\sigma^2_{\rm fw}(T-\tau)} - k(T-\tau),
\ee
and the corresponding FPE,
\be
\partial_\tau P(x,\tau) =  \frac{1}{2}\partial_x\left[ g^2(T-\tau) \partial_x  + k_{\rm bw}(\tau) x \right] P(x,\tau).
\ee
The solution is of the same form as for the forward process, see Eq.\ (\ref{eq:sigma2}),
\be
\label{eq:sigma2bw}
\sigma^2_{\rm bw}(\tau) =  \sigma^2_{0, \rm bw}f^2(\tau,0) + \Xi(\tau),
\ee
but with a more involved coefficient $k_{\rm bw}(\tau)$ and 
\be
f(\tau,s) = e^{-\half\int_s^\tau ds'\, k_{\rm bw}(s')},
\qqquad
\Xi(\tau) =  \int_0^\tau ds\, f^2(\tau,s)g^2(T-s).
\ee

We now consider two special cases:
\begin{enumerate}
\item Variance-expanding scheme without a drift term. We take $k(t)=0, g(t)=\sigma^{t/T}$, such that $f(t,s)=1$ and 
\be
\sigma^2_{\rm fw}(t) = \sigma^2_{\rm target} + \int_0^t ds\, g^2(s) = \sigma^2_{\rm target} + \frac{T}{2\log \sigma}\left[ \sigma^{2t/T}-1\right].
\ee
For $t/T\ll 1$, this reads $\sigma^2_{\rm fw}(t) = \sigma^2_{\rm target} + t$, a linear increase in time.
The memory of the initial distribution is suppressed at $t=T$, provided that $D\sigma^2\gg \sigma^2_{\rm target}$ and large, such that
\be
\sigma^2_{\rm fw}(T) \approx D\sigma^2,\qquad \mbox{with} \qquad
D = \frac{g^2(t)}{dg^2(t)/dt}  = \frac{T}{\log\sigma^2}.
\ee
The solution (\ref{eq:sigma2bw}) of the backward process has the somewhat cumbersome coefficients
\begin{align}
f(\tau,0)  & =  \frac{D \left( \sigma^{2(1-\tau/T)} -1\right) + \sigma^2_{\rm target}}{D \left(\sigma^{2}-1\right) + \sigma^2_{\rm target} }, \\
\Xi(\tau) & = D \sigma^{2(1-\tau/T)} \left(  \sigma^{2\tau/T} -1 \right)  f(\tau,0).
\end{align}
Since $f(0,0)=1$ and $\Xi(0)=0$, the solution satisfies the correct initial condition. At the end of the backward process, we find
\be
f(T,0) =  \frac{ \sigma^2_{\rm target}}{D \left( \sigma^2-1\right)  + \sigma^2_{\rm target} }, 
\qqquad
\Xi(T) =   D  \left(  \sigma^2 -1 \right) f(T,0).
\ee
In the same limit as above, $D\sigma^2\gg \sigma^2_{\rm target}$ and large, one finds that 
\begin{align}
 f(T,0) & =  0 +  \frac{ \sigma^2_{\rm target}}{D \left(\sigma^2-1\right)} +\ldots, \\
\Xi(T) & =   \sigma^2_{\rm target} \left[ 1 -   \frac{ \sigma^2_{\rm target}}{D \left(\sigma^2-1\right)} +\ldots \right],
\end{align}
such that the desired outcome is indeed obtained, $\sigma^2_{\rm bw}(T) \approx  \sigma^2_{\rm target}$.

Writing $\tau=T-\eps$, the approach to the value at $\tau=T$ can be analysed, using
\begin{align}
    f(T-\eps,0) & = f(T,0)\left[  1+\frac{\eps}{\sigma_{\rm target}^2} +\cO(\eps^2) \right], \\
    \Xi(T-\eps) & = \Xi(T) \frac{f(T-\eps,0)}{f(T,0)} \left[ 1-\frac{\eps^2}{D} +\cO(\eps^3)\right],    
\end{align}
which yields
\be
\sigma_{\rm bw}^2(T-\eps) = \sigma_{\rm bw}^2(T)\left[ 1+ \frac{\Xi(T)+2f^2(T,0)\sigma_{0,\rm bw}^2}{\Xi(T)+f^2(T,0)\sigma_{0,\rm bw}^2} \frac{\eps}{\sigma_{\rm target}^2} +\cO(\eps^2) \right].
\ee
Using again that $D\sigma^2\gg \sigma^2_{\rm target}$, this reduces to 
\be
\sigma_{\rm bw}^2(T-\eps) \approx \sigma_{\rm bw}^2(T) +  \frac{\sigma_{\rm bw}^2(T)}{\sigma_{\rm target}^2} \eps +\cO(\eps^2)
\approx
\sigma_{\rm bw}^2(T) + \eps +\cO(\eps^2),
\ee
i.e.\ the target value is approached linearly with slope 1, as expected from the forward process at early times.

\item Denoising diffusion probabilistic models (DDPMs). We take $k(t)=g^2(t)$, which incorporates various examples of DDPMs in the continuous-time limit. The FPE simplifies considerably, 
\be
\partial_t P(x,t) = \half g^2(t) \partial_x\left( \partial_x  + x \right) P(x,t),
\ee
which suggests to redefine time as
\be
u(t) = \int_0^t ds\, g^2(s),
\ee
such that 
\be
\partial_u P(x,u) =  \frac{1}{2}\partial_x\left(  \partial_x  + x \right) P(x,u).
\ee
With 
\be
f(t,s) = e^{-\half u(t)+\half u(s)}, \qqquad \Xi(t) = 1-f^2(t,0),
\ee
the variance during the forward process is given by
\be
\sigma^2_{\rm fw}(t) = 1 + e^{-u(t)}\left(  \sigma^2_{\rm target} -1\right),
\ee
with $\sigma^2_{\rm fw}(0) = \sigma^2_{\rm target}$.
At the end of the forward process, the variance is unity, up to exponentially suppressed terms. 
  
For the backward process, we introduce
\be
  v(\tau) = \int_0^\tau ds\, g^2(T-s) = u(T)-u(T-\tau).
\ee
We then find 
\begin{align}
f(\tau,s) & = e^{\half v(s)- \half v(\tau)} \frac{e^{v(T)}+c e^{v(\tau)}}{e^{v(T)}+c e^{v(s)}}, \\
\Xi(\tau) & = \left(1-e^{-v(\tau)}\right) \frac{1+c e^{v(\tau) - v(T)}}{1+ce^{-v(T)}},
\end{align}
with $c=\sigma^2_{\rm target} -1$. Inserting these in Eq.~(\ref{eq:sigma2bw}) yields the solution of the backward process. 
At the end of the backward process, we obtain
\be
f(T,0)  =  \frac{e^{-\half v(T)}}{1+c e^{-v(T)}} \sigma^2_{\rm target} \to 0, 
\qquad
\Xi(T)  = \frac{1-e^{-v(T)}}{1+ce^{-v(T)}} \sigma^2_{\rm target} \to \sigma^2_{\rm target},
\ee
which implies that the desired outcome is again obtained, $\sigma^2_{\rm bw}(T) \approx  \sigma^2_{\rm target}$, up to exponentially suppressed terms. Note that this is independent of the choice of $g(t)$. 

\end{enumerate}

We conclude that in the case of a Gaussian target distribution both schemes lead to the correct result, by explicit computation. Noticeably, the manner in which this is achieved is quite different: in the scheme without a drift, the variance at the end of the forward process should become large, and in particular much larger than target variance. Corrections are suppressed as $\sigma^2_{\rm target}/\sigma^2$, where $\sigma^2$ is the strength of the noise at $t=T$. In the scheme with a drift this requirement is not needed: the variance at the end of the forward process becomes unity and deviations are suppressed exponentially.

\providecommand{\href}[2]{#2}\begingroup\raggedright\endgroup


\end{document}